**Title :**

AC magnetic behavior of large grain magneto-resistive $La_{0.78}Ca_{0.22}Mn_{0.90}O_x$ materials

**Authors :**

Ph. Vanderbemden [a], B. Vertruyen [b], A. Rulmont [b], R. Cloots [b], G. Dhalenne [c], and M. Ausloos [d]

[a] SUPRATECS and Department of Electrical Engineering and Computer Science (Montefiore Institute) B28, University of Liège, Sart-Tilman, B-4000 Liège, Belgium

[b] SUPRATECS and Chemistry Institute B6, University of Liège, Sart-Tilman, B-4000 Liège, Belgium

[c] Laboratoire de Physico-Chimie de l'Etat Solide, Université Paris-Sud, 91405 Orsay Cedex, France

[d] SUPRATECS and Physics Institute B5, University of Liège, Sart-Tilman, B-4000 Liège, Belgium

**Corresponding author :**

Dr. Philippe Vanderbemden
University of Liège, SUPRATECS,
Department of Electrical Engineering and Computer Science (Montefiore Institute) B28,
Sart-Tilman,
B-4000 Liège,
Belgium.
Phone : +32 4366 2674
Fax : +32 4366 2877
e-mail : Philippe.Vanderbemden@ulg.ac.be

**Abstract**

We report a detailed set of AC magnetic measurements carried out on bulk large grain La-Ca-Mn-O samples extracted from a floating zone method-grown rod. Three samples with $La_{0.78}Ca_{0.22}Mn_{0.90}O_x$ stoichiometry but differing in their microstructure were investigated by electrical resistivity and AC susceptibility measurements: (i) a single grain sample, (ii) a sample containing two grains and (iii) a polycrystalline sample. We show that the superimposition of DC magnetic fields during AC magnetic susceptibility measurements is an efficient way for characterizing the magnetic transition of samples with different microstructures. Whereas both single grain and polycrystalline samples display a single susceptibility peak, an additional kink structure is observed in the case of the double grain sample. The temperature dependence of the AC susceptibility measured with superimposed DC magnetic fields is analyzed in the framework of second-order phase transition ideas. The relations between the critical exponents ($\beta + \gamma \sim 1.5$, $\delta \sim 2.5$) are found to be close to those of the mean-field model for all samples. This is attributed to the disordering caused by unoccupied Mn sites.

# I. INTRODUCTION

The interest in manganese perovskite compounds of the $Ln_{1-x}A_xMnO_{3+/-d}$ family (where Ln is a large lanthanide and A generally an alkaline-earth) has been recently revived after the discovery of CMR (Colossal MagnetoResistance) properties in some of these materials :[1-3] the electrical resistivity, exhibiting a maximum at a given temperature $T_{MI}$ corresponding to a metal-insulator transition, is drastically suppressed under the application of a magnetic field. These materials are also characterized by a ferromagnetic-paramagnetic transition at a temperature $T_C$ close to $T_{MI}$ noticeable in various physical properties.[4]

More generally, the research activity on CMR materials brings out underlying fundamental aspects which are of great interest for the physics of highly correlated electron systems.[2] The physical properties of these compounds are influenced by several parameters. The two most meaningful ones are the $Mn^{4+}/Mn^{3+}$ ratio (i.e. the charge carrier density)[5] and the Mn-O-Mn bond angle, which affects the orbital overlapping between neighboring ions.[6]

Beside these intrinsic parameters, the microstructure of these materials was shown to influence strongly their electrical transport properties, as proved by comparative studies of thin films, bulk ceramics and single crystals.[7-16] In single crystals and epitaxial thin films, the magnetoresistance is quite large and concentrated in the vicinity of the transition temperature $T_C$ whereas in polycrystalline materials – either bulk ceramics or thin films – a significant magnetoresistance is displayed at low fields for all temperatures below $T_C$.

Unlike the *transport* properties, the *magnetic* properties of polycrystalline CMR materials were shown to be weakly influenced by their microstructure.[15,17] To our knowledge however, no systematic study of magnetic properties has been performed on bulk CMR material containing either one single grain or two grains separated by a single grain boundary. In the present study we report and discuss AC magnetic susceptibility measurements carried out on such large grain samples, with the emphasis placed on the study of magnetic fluctuations around the transition temperature. In a previous work[18] we have reported the characterization of these samples through electrical resistivity and DC magnetization measurements.

AC magnetic susceptibility measurements have been widely used for characterizing the magnetic transitions occurring in various materials,[19,20] including CMR materials.[17,21-23] However, in such complex materials as manganites, the actual magnetic structure often results from the competition between several magnetic states of similar ground-state energy (ferromagnetic, antiferromagnetic, charge-ordered,…).[24,25] This can even lead to so-called phase-separated materials, where two magnetic phases coexist in a single crystal.[26,27] Consequently, the physical mechanisms governing the magnetic response cannot always be distinctly sorted out through an AC susceptibility experiment. As an example, a frequency dependence of the AC susceptibility can be due to either an intrinsic spin glass behavior or to extrinsic phenomena such as domain wall pinning.[28] Depending on the sample homogeneity, the magnetic transition may also occur over a wide temperature range.[29] Therefore the analysis of an AC magnetic response can be difficult and sometimes unconclusive.

When the temperature dependence of the AC susceptibility is measured in presence of DC field, the low temperature ($T < T_C$) signal contribution due to the physical mechanisms mentioned above is progressively driven to saturation, allowing the emergence of a universal signal contribution arising from critical fluctuations.[23] As a result, a sharp peak in the in-phase AC susceptibility $\chi'(T)$ can be clearly identified near $T_C$. A comprehensive review of the theoretical and experimental aspects relative to the origin of this peak and its field dependence has been carried out by Williams,[23] within the classical framework of a second order paramagnetic / ferromagnetic transition theory. The presence of the peak was shown experimentally in dilute magnetic systems such as AuFe,[30] PdMn,[31] and amorphous ferromagnetic alloys.[32] More recently, a similar behavior was also depicted in CMR ceramics such as $La_{0.67}Pb_{0.33}MnO_3$ [33], $La_{0.67}Ca_{0.33}MnO_3$ [17,34] and $La_{1-x}Mg_xMnO_3$.[35-37] It should be emphasized that all these works refer to polycrystalline samples containing grains whose size is typically smaller than ~ 50 µm. The present study is concerned with the comparison of the magnetic properties of bulk La-Ca-Mn-O samples containing either (i) one single large grain or (ii) two large grains separated by a single grain boundary. The results are compared to those obtained in polycrystalline samples of the same material.

## II. EXPERIMENTAL

### A. Synthesis process

A 30-mm long 4-mm diameter cylindrical rod of calcium-doped lanthanum manganate (LCMO) was grown by the floating zone method. The details of the synthesis procedure as well as specific growth features have been described in a previous paper.[18] The material microstructure was examined by polarized light optical microscopy (Olympus AH3-UMA). The observation of the rod cross section at several locations between both ends shows that the mean grain size progressively increases and finally reaches ~ 1 mm$^3$ near the far end of the rod, as sketched in Fig. 1. A polarized light micrography of a cross-section in the far end of the rod is also shown in Fig. 1, revealing the presence of only three large grains. Three bar-shaped samples of typical 0.1 x 0.2 x 0.8 mm$^3$ size were carefully excised from the rod using a wire saw. Their microstructures are single grain (SG), double grain (DG), and polygranular (PG).

Energy Dispersive X-ray (EDX) analysis (Oxford Link Pentafet) of each sample showed an homogeneous chemical composition, within the uncertainty of the EDX method. However, this chemical composition was found to differ from the nominal stoichiometry ($La_{0.7}Ca_{0.3}MnO_3$). This phenomenon is due (i) to the manganese loss by vaporization during the growth of the rod and (ii) to the low value of the calcium distribution coefficient between solid and liquid phases.[38] More precisely, the cationic composition determined by EDX turns out to be $La_{0.78}Ca_{0.22}Mn_{0.90}O_x$. Moreover, the knowledge of the density (6.07, measured by the Archimedes' method) and the cell volume (233.9 Å$^3$, refined from XRD data in the Pbnm space group, with the FULLPROF software) has enabled us to calculate the molar mass. The oxygen content could thus be estimated, yielding a chemical composition close to $La_{0.78}Ca_{0.22}Mn_{0.90}O_{2.94}$. The theoretical number of Bohr magnetons estimated for such a chemical content (3.20 $\mu_B$) is in good agreement with the experimental value (3.17 $\mu_B$) determined by measuring the saturation magnetization at T = 50 K and $\mu_0 H$ = 5 T.[18]

### B. Physical measurements

DC magnetic moment measurements at several temperatures were carried out in a Quantum Design Physical Property Measurement System (PPMS), using an extraction method. AC magnetic susceptibility measurements were performed both in a home-made susceptometer[39] and in the PPMS. Before each measuring sequence, the remnant field of the superconducting magnet was eliminated by applying a succession of decreasing fields in alternate directions.

Transport measurements were carried out on the three specimens using the conventional 4-point technique. Very small electrical contacts were achieved by attaching thin gold wires (33 μm diameter) to the samples using DuPont 6838 silver epoxy paste annealed in flowing $O_2$ for 5 minutes. In the sample containing two grains (DG), the electrical contacts were placed across the single grain boundary. The electrical resistance vs. temperature R(T) curves measured under applied DC magnetic fields ranging from 0 to 1 T were recorded in the Quantum Design PPMS between 20 K and room temperature.

## III. RESULTS

### A. ELECTRICAL RESISTIVITY

The temperature dependence of the electrical resistivity of the three La-Ca-Mn-O samples is shown in Fig. 2. The data are measured with a 1 mA injection current parallel to the long axis of each sample. All samples display the overall characteristics of a transition from a low temperature metallic-like state ($d\rho/dT > 0$) to a high temperature insulator-like behavior ($d\rho/dT < 0$). Both the single grain (SG) and double grain (DG) samples display a sharp peak at $T = T_p \sim 196$ K and very similar electrical resistivity values at temperatures $T > T_p$. Their behavior markedly differs for $T < T_p$ : the resistivity of the single grain sample (SG) is significantly lower than the resistivity measured across the grain boundary in the sample containing two grains (DG). At $T = 20$ K, the electrical resistivity values for SG and DG samples are 0.57 and 2.2 μΩ.m respectively. The use of a semi-log scale in Fig.2 allows us to compare qualitatively the ρ($T$) curves of both samples for $T < T_p$ : the double grain sample exhibits a slight shoulder structure whereas the data measured for the single grain do not display any inflexion point. The presence of such a shoulder in the resistivity curve is the signature of the presence of a grain boundary in the DG sample,[18] while the data measured on the SG sample are similar to those measured on LCMO single crystals.[7,25,30,41] The clear differences in the resistivity behavior of the SG and the DG samples do also confirm, *a posteriori*, that no "unseen" grain boundary is present in the "single grain".

The polygranular sample (PG) is characterized by much higher electrical resistivity values than the SG and DG samples. For $T > T_p$, the resistivity of PG lies one order of magnitude above that of SG and of DG. The PG resistivity peak around $T \approx T_p$ is quite smooth but perceptibly emerges from the large resistivity signal occurring at $T < T_p$. At $T = 20$ K, the electrical resistivity of the PG sample is 2500 μΩ.m, *i.e.* three orders of magnitude above the resistivity of the DG sample. All these characteristics are consistent with the polycrystalline nature of the PG sample containing a significant number of grain boundaries, which inhibit the current flow and are thus responsible for the higher resistivity values.

The transition temperature $T_{MI}$ of each of the three samples was determined by locating the main inflexion point of ρ($T$), yielding values of 190.2 K, 188.5 K and 192.5 K for the SG, DG, and PG samples respectively. It should be noticed however that the sharpest of the three resistive transitions, i.e. that of the SG sample, is expected to be the most appropriate for getting an accurate $T_{MI}$ determination. These $T_{MI}$ values are in agreement with those reported in the literature for similar chemical composition.[40]

B. AC SUSCEPTIBILITY

The temperature dependence of the AC susceptibility of the three samples was first measured for a 1 mT and 1 kHz applied AC magnetic field *without* bias DC magnetic field. All reported measurements (Fig. 3) were carried out in zero-field cooling, but no noticeable difference was observed with respect to the field cooled procedure, within experimental uncertainty. On lowering the temperature, the susceptibility increases rapidly when the system undergoes the metal-insulator transition at $T = T_C$, becoming nearly temperature independent below $T_C$. A careful examination of the data shows that the susceptibility passes through a maximum (the so-called Hopkinson peak[43]), and then slowly decreases with a very small (d$\chi$'/d$T$) value. The behavior is in good overall agreement with existing measurements on other CMR samples[17,21-23] and displays the characteristics of a classical paramagnetic – ferromagnetic phase transition.[42,43] The magnetic transition is somewhat sharper for the single grain sample SG than for the DG and PG samples. The rather small (d$\chi$'/d$T$) observed for the three samples at $T < T_C$ strongly suggests that the $\chi$'($T$) dependence is bounded to some value determined by the sample geometry. Using the classical notations, the internal magnetic field $H_i$ is given by $H_a - D.M$, where $H_a$ and $M$ respectively denote the applied field and the sample magnetization; $D$ is the demagnetization factor ($0 < D < 1$). For materials exhibiting a high susceptibility ($M / H_i$), the measured *apparent* susceptibility ($M / H_a$) is limited to a maximum value roughly given by 1/$D$. This limit is fixed by the sample dimensions and is therefore temperature independent, as observed in Fig. 3. Using the data of Fig. 3, one can estimate the demagnetization factors of the SG, DG, PG samples to be respectively 0.16, 0.22 and 0.10, consistently with the values of 0.14, 0.19 and 0.09 estimated from the sample dimensions.[44] Hence the differences in the low temperature $\chi$' values for the three samples are caused by their geometry rather than by their microstructure. It is well-known that small-$D$ (i.e. long and thin) samples should be preferred for the study of magnetic properties but it was not possible to extract long specimens in the case of our quasi single-grain materials. Therefore these geometrical effects have to be taken into account in the present study.

The temperature dependence of the AC susceptibility was also measured under various superimposed DC magnetic fields. Both AC and DC magnetic fields were parallel to the long axis of the samples. The inset of Fig. 3 shows the typical evolution of the AC susceptibility in-phase component $\chi$' for the DG sample under increasing DC fields ranging from 0.1 to 0.3 T. A distinct peak appears around the transition temperature. Figure 4 focuses on the evolution of this $\chi$'(T) peak for larger DC bias fields, i.e. from 0.3 T to 1 T. The $\chi$' data plotted in Fig. 4 are corrected for demagnetization effects using the demagnetization factor of each specimen determined as above. All samples display the same overall behavior : when increasing the DC field amplitude, the maximum of $\chi$' shifts to higher temperatures, decreases in amplitude and is progressively smeared out, in agreement with the data reported previously for other ferromagnetic systems.[23]

Strikingly however, the DG sample displays a well-defined kink structure, which is not observed in the single grain (SG) and the polycrystalline (PG) samples. This can be clearly seen in the upper curve in Fig. 4b ($\mu_0H = 0.3$ T) : the main peak, located at $T_1 = 193.2$ K, is followed by a kink around $T_2 = 196.5$ K. On increasing the applied DC field amplitude, the behavior of this kink mimics that of the main peak. For applied magnetic fields exceeding ~ 0.8 T both peaks merge into one large bump. Measurements with a magnetic field perpendicular to the long axis of the DG sample ($D \sim 0.42$) display a behavior entirely similar to the one depicted in Fig. 4b but with slightly different peak temperatures.

In Fig. 5, we have plotted the main peak temperature $T_1$ of the DG sample as a function of the *internal* magnetic field $H_i$, calculated by the formula $H_i = H_a - D.M$, where $H_a$ is the applied magnetic field and $M$ is the DC magnetization value, carefully measured at each ($T_1$, $H_a$) point, i.e. at the peak temperature $T_1$ corresponding to the applied field $H_a$. As can be seen on the figure, the data sets collected for each DC field orientation follow one single curve when plotted as a function of the *internal* magnetic field. This indicates that there is no intrinsic anisotropy in the susceptibility behavior of the sample.

In summary, the results show that (i) the resistivity data confirm what could be expected from the microstructure of each sample, (ii) the resistive and magnetic transitions are the sharpest for the single grain sample SG, (iii) the zero-DC field AC susceptibility behavior is dominated by geometric effects, (iv) the DG sample displays a perceptible kink structure in the $\chi'(T)$ data measured under bias static magnetic fields. The details, differences and similarities in the AC magnetic properties of the three samples are the subject of Sect. IV.

**IV. DISCUSSION**

First of all, it should be noticed that the transition temperature of all studied samples lies around 190 K, which is lower than the transition temperature ($T_C \approx 260$ K) characteristic of the $La_{0.7}Ca_{0.3}MnO_3$ stoichiometry. In fact, the transition temperature lies between those measured for $La_{1-x}Ca_xMnO_3$ single crystals[40] with $x = 0.225$ and $x = 0.275$. This feature can be attributed to the actual chemical composition of the sample ($La_{0.78}Ca_{0.22}Mn_{0.90}O_{2.94}$), which displays Mn deficiency and lower Ca/La ratio with respect to the nominal composition (see Section IIA).

A. Critical fluctuations analysis

In ref. [17] dealing with the magnetic properties of a $La_{0.67}Ca_{0.33}MnO_3$ polycrystalline material, it was shown that the locus of the $\chi'(T)$ maxima measured for several DC fields defines a crossover line above which the magnetic response is thermally dominated, and below which the response is field dominated. In terms of the usual reduced fields and temperatures given by $h \sim H_i / T_C$ and $t = |T - T_C| / T_C$ and using the scaling law equation of state,[23] the product

$$h \; t_m^{-(\gamma+\beta)}$$

should be a constant. In this equation, $\gamma$ and $\beta$ are the critical exponents and $t_m$ denotes the reduced temperature at the $\chi'$ peak. This suggests that the peak temperature $T_p$ measured in all samples should fit a relationship given by

$$T_p = T_C + a.H_i^n,$$

with an exponent $n$ equal to $1/(\beta + \gamma)$. The fitting parameters obtained for the SG, DG and PG samples are listed in Table I. The DG sample case is illustrated in Fig. 5. The $(\beta + \gamma)$ values are seen to range between 1.39 and 1.61, closer to the mean-field approximation ($\beta + \gamma = 1.5$) than to the 3D Heisenberg prediction ($\beta + \gamma = 1.75$).[45] The procedure also allows us to determine precisely the critical temperature $T_C$ of each sample by extrapolating the results down to $H_i = 0$. The results, summarized in Table I, show a very good agreement between the "theoretical" magnetic $T_C$ and the corresponding $T_{MI}$ values deduced from the electrical transport measurements shown above (Fig. 2).

Similarly, scaling arguments[23] show that the amplitude of the peak susceptibility $\chi_m$ should follow a power law relationship as a function of the reduced internal field $h$

$$\chi_m \sim h^{(1/\delta)-1},$$

so written using the Widom equality[23] $\gamma = \beta(\delta-1)$. The $\delta$ values obtained by fitting the peak amplitude (corrected for demagnetizing effects) as a power law function of the internal field are listed in Table I. Notice that the $\delta$ exponent is determined without any other assumption on $\beta$, $\gamma$ or $T_C$. As can be seen, the $\delta$ values for the three samples (ranging between 2.42 and 2.67) are not consistent with the 3D Heisenberg ($\delta = 4.803$) predictions[46] but are rather close to the mean-field value ($\delta = 3$). In the particular case of the DG sample exhibiting the kink structure, the fitting procedure was also carried out using the ($T^*$, $\chi'^*$) points resulting from the intersection of lines extrapolated from outside the temperature window containing the peak and the kink, as shown by the solid lines in Fig. 4b. However, such a procedure does not significantly modify the results; it leads to $\delta = 2.52$, a value close to $\delta = 2.54$ obtained by locating the true maximum of the experimental data without any curve fitting.

The discrepancy between the experimental data and the theoretical predictions can be discussed as follows. Some authors[35] have shown that the paramagnetic to ferromagnetic transition in the case of the $La_{1-x}Ca_xMnO_3$ system might be either first order ($x \sim 0.3$) or second order ($x \sim 0.2$). Since the actual transition temperature of the presently investigated material is close to that of $La_{1-x}Ca_xMnO_3$ with $x \sim 0.25$, some ambiguity can be expected. It has sometimes been suggested that the nature of the magnetic transition in $La_{0.67}Ca_{0.33}MnO_3$ differs from that of other CMR materials[41] and that a description of this compound could be made in terms of percolation theory for phase-separated clusters.[47] In the case of the present samples, it clearly appears that the microstructure has little effect on the critical exponents which are found to be close to mean-field values. We consider that this feature is related to the Mn deficiency in the actual material stoichiometry : the unoccupied Mn sites undoubtedly lead to an increase of the relative impact of long distance interactions in the compound. This, in turn, suggests a decreasing correlation range of the fluctuations, which means that the mean-field approximation might be appropriate for describing the magnetic fluctuations occurring in the samples.[48]

## B. Origin of the kink (DG sample)

In order to investigate the reason for the peculiar behavior of the DG sample, careful resistivity measurements were carried out in the vicinity of the temperature and magnetic field line shown in Fig. 5, but no noticeable singularity could be detected. In addition, the results displayed in Fig. 5 show that the sample geometry does not affect the material behavior, since data points measured with a magnetic field parallel ($D = 0.22$) or perpendicular ($D = 0.42$) to the long axis of the sample follow one unique line. We also emphasize that all peak amplitudes plotted in Fig. 4 lie well below the demagnetization limit ($1/D$). Therefore it can be concluded that the $D$ factor – despite its rather high value – is not a relevant parameter for the analysis of the phenomenon.

In the literature, the only occurrence of a double bump structure for $\chi'$ was reported for some dilute magnetic systems such as PdMn alloys.[31] As the Mn concentration increases from 3% to 5%, a secondary peak appears at some temperature below the main peak characterizing the critical fluctuations. Both peak amplitudes were shown to decrease rapidly with increasing DC magnetic field, but, unlike the behavior depicted in Fig. 4b, the low-$T$ peak was shown to be shifted towards *lower* temperatures as the field increases.[31] Such a behavior is similar to what is observed in spin glasses, but this is obviously not the behavior observed here.

Based on these considerations, we propose that the kink structure in the DG sample can be attributed to a slight difference, i.e. ~ 3K, between the critical temperatures of the two constitutive grains. This feature may be caused by a small difference in their respective stoichiometry, not perceptible through the resolution of the EDX analysis. This interpretation is consistent with the fact that no kink could be observed, neither for the single grain nor the polycrystalline sample. In the case of the single grain sample, the stoichiometry is expected to be uniform, resulting in a unique critical temperature as observed in Fig. 4a. In the case of the polycrystalline sample, the numerous grains might still have slightly different stoichiometries – and thus slightly different $T_C$'s – but the overall magnetic properties of the sample are averaged on a length scale which is at least one order of magnitude larger than the average grain size. The $T_C$ distribution is thus expected to be completely rounded off and only one well-defined large peak appears, as shown in Fig. 4c.

The results obtained for the DG sample put into evidence that specific features may sometimes be observed when samples have a size comparable to the grain size itself. In that respect, AC susceptibility measurements in the presence of DC fields is a powerful tool to reveal small $T_C$ inhomogeneities within the material and assess the sample quality. It is also worth emphasizing that, in the case of polycrystalline materials, the presence of one single peak in the AC susceptibility vs. temperature curve is not a strict proof of the sample homogeneity. The peak may indeed result from the superposition of several peaks very close to each other, reflecting the $T_C$ distribution in the sample.

## V. CONCLUSIONS

We have examined the properties of three magnetoresistive La-Ca-Mn-O samples ($La_{0.78}Ca_{0.22}Mn_{0.90}O_{2.94}$) extracted from a rod grown by the floating zone method. The samples are characterized by different microstructures and contained either (i) one single grain, (ii) two large grains, or (iii) several small grains. The material chemical composition was determined to be homogeneous within the uncertainty of the EDX method. The quality of the LCMO bulk material was confirmed by both resistivity and AC susceptibility measurements. Superimposing a DC field on the AC driving field led to the appearance of a maximum in $\chi'$, whose field and temperature dependence is consistent with the description of a second-order magnetic transition. The critical exponent values ($\beta + \gamma \sim 1.5$, $\delta \sim 2.5$) were found to be independent of the microstructure. These relations between critical exponents are close to those of the mean-field approximation. This can be understood from the actual stoichiometry of the investigated samples in which disordered unoccupied Mn sites cause a shortening of the fluctuation correlation length.

In the sample containing two grains, a noticeable kink structure in the AC susceptibility was observed. This phenomenon was interpreted as being the signature of a small difference between the critical temperatures of the adjacent grains. Such results emphasize the usefulness of AC magnetic measurements in the presence of DC fields in order to bring out small $T_C$ variations within the sample. The kink feature was observed neither in the single grain material nor in the polycrystalline sample. In this latter case, the properties are expected to be averaged over several grains and the data display only one peak, in spite of possible sample inhomogeneities. Therefore we can conclude that considerable caution needs to be taken when studying magnetic measurements on non-homogeneous samples.

**ACKNOWLEDGMENTS**

Ph. V. and B.V. are grateful to the F.N.R.S. for a Scientific research worker grant and for a Research fellow grant respectively. We also would like to thank Prof. H.W. Vanderschueren for allowing us to use the M.I.EL. laboratory facilities. B.V. thanks Prof. Revcolevschi and Prof. Berthet for welcoming her at the Laboratoire de Physico-Chimie de l'Etat Solide.

Table I : Comparison of resistive transition temperature $T_{MI}$, critical temperature $T_C$ and critical exponent (β + γ) and δ values determined for the three samples. For "DG (extrapolation)", see the procedure described in the text.

| Sample | $T_{MI}$ (K) | $T_C$ (K) | β + γ | δ |
|---|---|---|---|---|
| **SG** | 190.2 | 190.0 | 1.61 | 2.42 |
| **DG** | 188.5 | 189.2 | 1.49 | 2.54 |
| **DG (extrapolation)** | --- | 190.5 | ---- | 2.52 |
| **PG** | 192.5 | 192.3 | 1.39 | 2.67 |

**Figures captions**

Fig. 1. (Left) schematic diagram of the far-end of the La-Ca-Mn-O rod showing the locations where three samples were extracted : SG = single grain, DG = double grain, PG = polygranular sample. (Right) optical polarized-light micrography of a cross section in the far end of the rod.

Fig. 2. Comparison of the electrical resistivity vs. temperature curves measured on the single grain (SG), the double grain (DG) and the polycrystalline (PG) samples.

Fig. 3. Temperature dependence of the real component ($\chi'$) of the AC susceptibility measured on the single grain (SG), the double grain (DG) and the polycrystalline (PG) samples. Inset : evolution of the $\chi'(T)$ curves of the DG sample under several superimposed DC magnetic fields ranging from 0 to 0.3 T.

Fig. 4. Real component ($\chi'$) of the AC susceptibility measured on the single grain (SG), the double grain (DG) and the polycrystalline (PG) samples for several superimposed DC magnetic fields ranging from 0.3 to 1 T, with 0.1 T steps. The lines in (b) represent data extrapolated from outside the temperature window containing the two peaks.

Fig. 5. Comparison of the in-phase AC susceptibility peak temperature vs. internal magnetic field measured for the applied magnetic field applied either parallel (white symbols) or perpendicular (black symbols) to the long axis of the double grain (DG) sample. Both sets of data are fitted by the same law (black line).

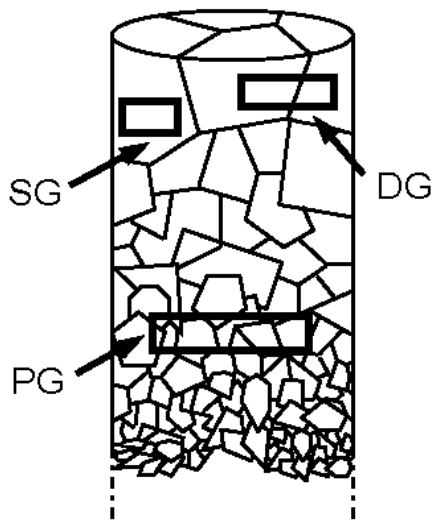 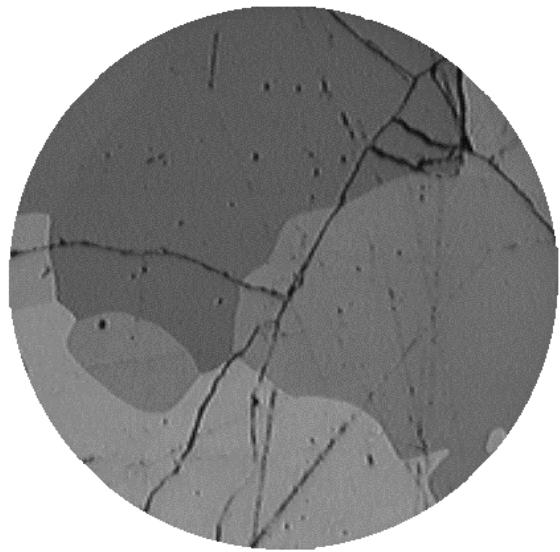

**Figure 1**

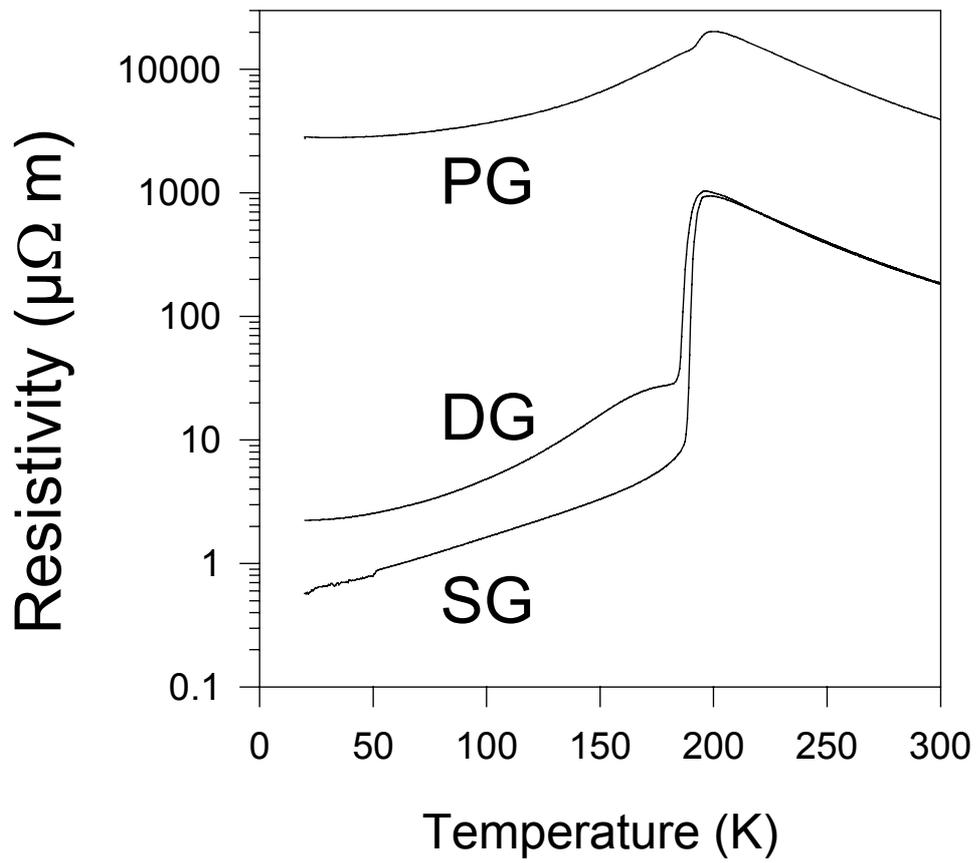

Figure 2

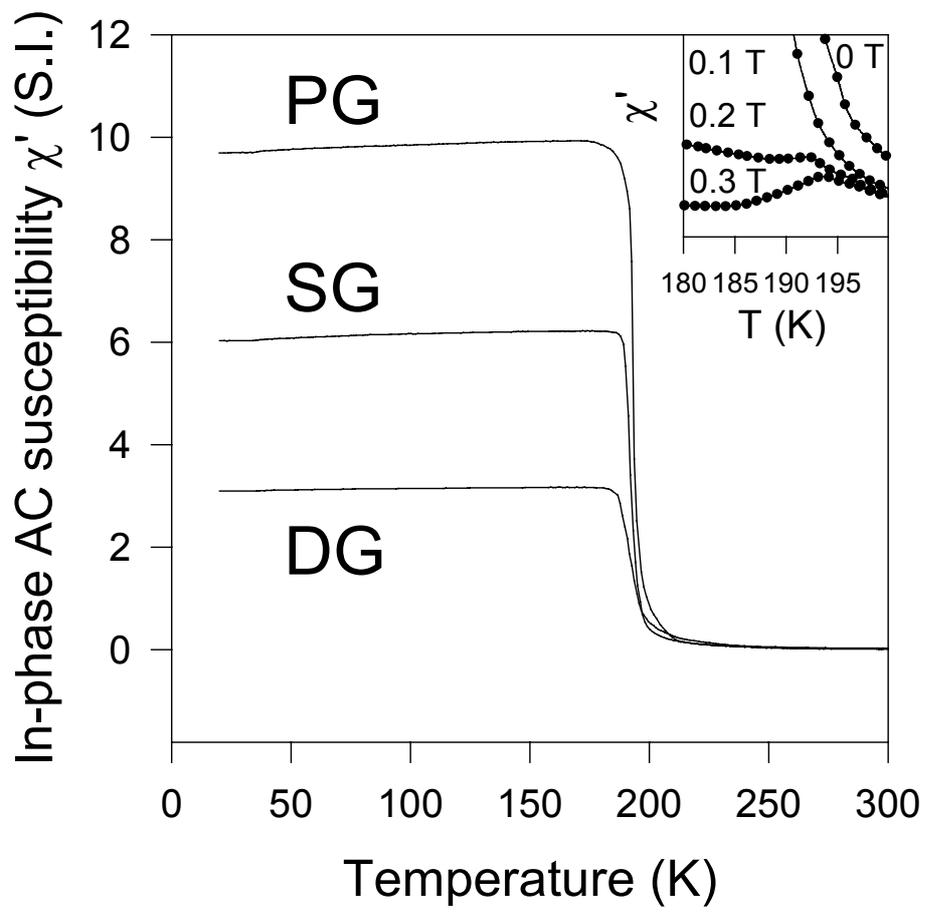

**Figure 3**

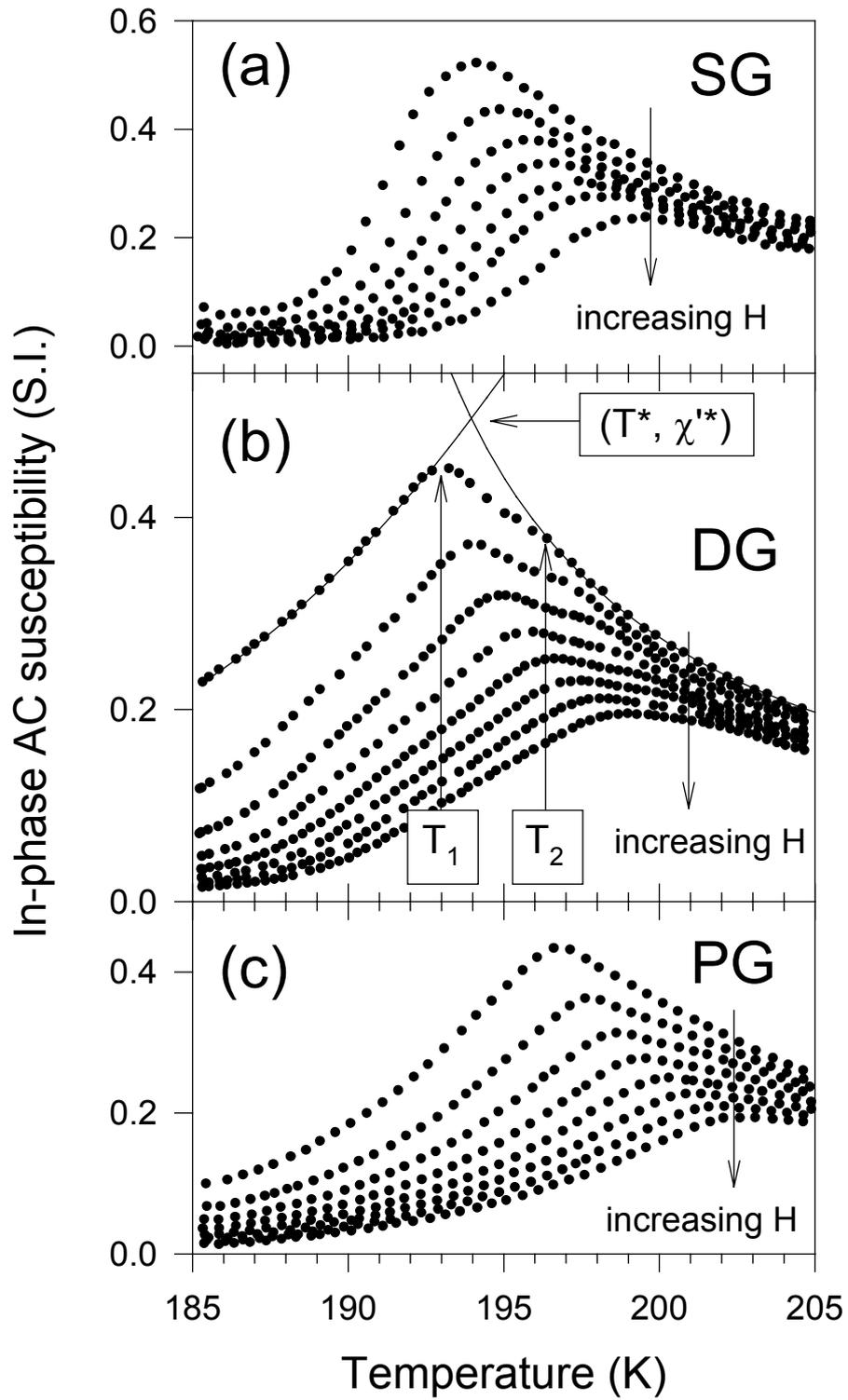

**Figure 4**

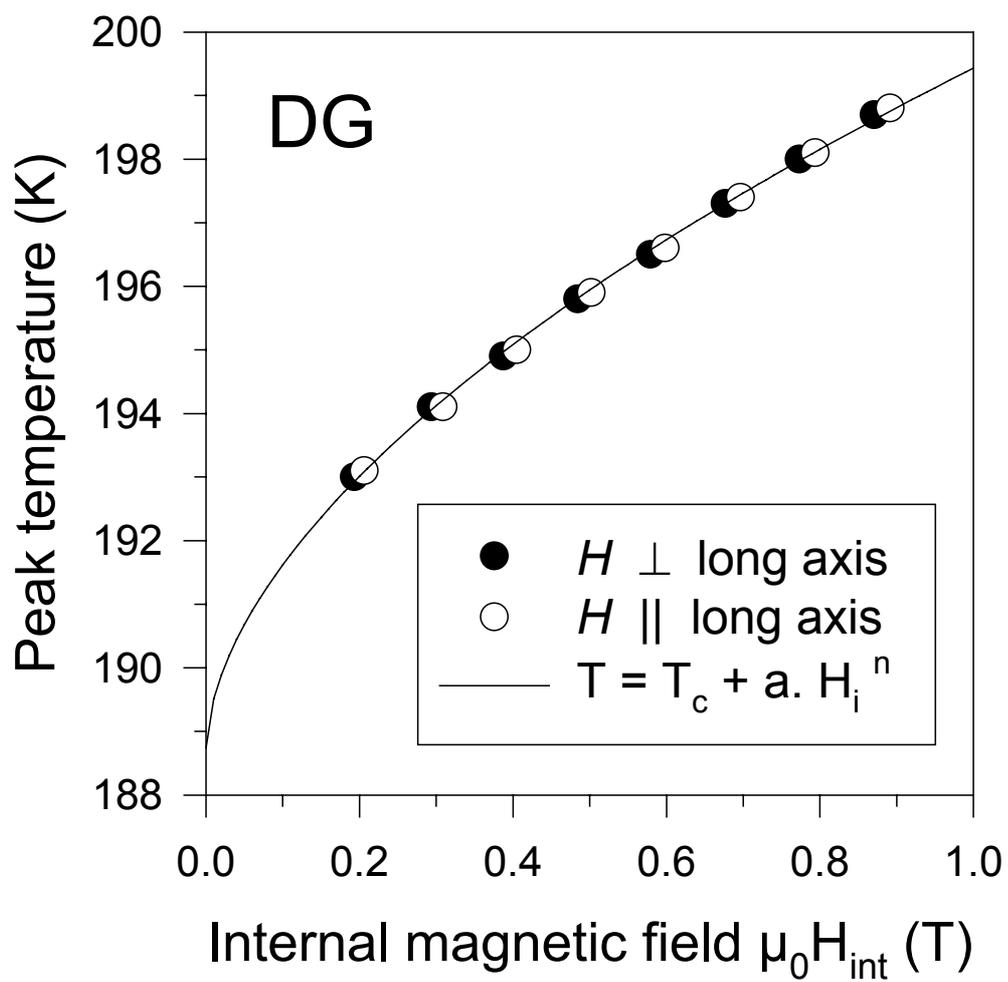

**Figure 5**